# Study of ITG modes in RFX-mod using TRB code

F. Sattin[1], X. Garbet[2], S.C. Guo[1]

[1] *Consorzio RFX, Associazione Euratom-ENEA per la fusione, Padova, Italy*
[2] *CEA, IRFM, F-13108 Saint Paul Lez Durance, France.*

**Abstract**

We present here a study about the stability of Ion-Temperature-Gradient drift turbulence in the Quasi-Single-Helicity regime of RFX-mod Reversed Field Pinch (RFP) using the TRB fluid electrostatic turbulence code. Our results suggest that present-days RFP plasmas are marginally stable against this kind of turbulence. The onset of the instability may be envisaged for close future regimes, in the presence of hotter plasmas with sharper internal transport barriers.

## 1. Introduction

Present-days Reversed Field Pinches (RFPs), thanks to sensible improvements in the feedback control of the magnetic boundaries, are noticeably less prone to large-scale magnetic chaos. The current paradigm for them is the Quasi-Single-Helicity (QSH) configuration, in which the total magnetic field is made of an axisymmetric component plus just one single helical mode, with poloidal number $m = 1$ and toroidal number $n^* \approx 2\times$(Major radius)/(Minor radius) (plus, of course, its harmonics); the presence of at least one non axisymmetric mode being required by physical constraints (the RFP counterpart of Cowling's theorem). Chaotic wandering of magnetic field lines in the core is suppressed, and a region of quite well-preserved magnetic surfaces emerges. As a consequence, fast radial transport due to parallel motion along magnetic field lines is strongly abated and a hot central region is observed, particularly evident in high-current experiments [1]. Two categories of helical states are observed, depending on the relative amplitude of the dominant mode normalized to axisymmetric component: (I) at relatively small amplitudes, the axysimmetric magnetic axis and the secondary magnetic axis coexist. The helical structure is bounded by a magnetic separatrix. In these cases, a localized temperature peak appears, associated with the magnetic island and of moderate extent (10-20% of the major diameter); (II) When the dominant mode grows up to the point where its axis becomes the main magnetic axis (about 4% of the edge magnetic field), the original symmetric magnetic axis ceases to exist. The high-temperature region is now considerably widened[1]. These helical states (dubbed SHAx, for Single Helical Axis) are expected to be resilient to the remaining magnetic chaos, and therefore particularly appealing for confinement [3]. Due to their intrinsic importance and since these states appear naturally at high current operations, which are the operating conditions of greater interest, we will refer only to SHAx's in the rest of the paper.

The mitigation of magnetic chaos opens the way for other physics to be observable and maybe become even dominant. In particular, the existence of strong gradients at the boundaries of the warmer region may trigger drift instabilities which act as damping mechanisms to further increase of $\nabla T$. In Tokamaks, one of the most important core drift instabilities is the Ion Temperature Gradient (ITG) mode. In this work we will carry out an analysis for such instability in RFX-mod device, currently the largest RFP experiment worldwide (minor radius 0.46 m, major radius 2 m, maximum plasma current 2 MA) [4], and the only one—up to date—where SHAx states have been observed.

An earlier gyrokinetic semi-analytical study provided evidence that the RFP is—by virtue of its short connection length—fairly resilient to ITGs but that—on the other hand—once triggered, these modes might have fairly large growth rates because of the strong magnetic curvature drift [5]. Present-days RFPs are thought to operate mostly away from the onset of instability, but in SHAx discharges featuring a particularly relevant increase of the central temperature, the interface zone between the warm core and the colder edge could actually be close to marginal stability.

Due to the traditional dominance of large-scale instabilities in RFP research, there is not abundancy of tools designed for studying microinstabilities. As far as RFX-mod is concerned, two pre-existing codes have been adapted to deal with the peculiarities of the RFP geometry. One is GS2: a powerful gyrokinetic code for the study of low-frequency modes [6]. The second, of relevance here, is the fluid code TRB, developed to handle with electrostatic drift instabilities in Tokamak geometries [7]. Both codes come with advantages and disadvantages, depending upon the exact target addressed. GS2, accounting for kinetic as well as electromagnetic effects, provides more precise

[1] Recently, signatures of the same phenomenology have been identified also from density and flow measurements [2]

calculations. On the other hand, the computational burden is such to limit the use of the code both spatially (GS2 works in flux-tube geometry) and temporally. Conversely, TRB allows global simulations extended over long times. A really convenient feature of TRB, here, is its capability of including within its fluid formalism a kinetic effect as the Landau Damping (LD). Fluid models are convenient tools for numerical computations, since they diminish the effective dimensionality of the systems studied by neglecting the velocity degrees of freedom. Their drawback is that in low-collisionality high-temperature plasmas departures from Maxwellian distribution could be relevant and should be properly accounted for. Neglect in doing so results in overestimating turbulent fluxes, as shown by Dimits *et al* [8]. This general result was confirmed by Guo for RFPs in [5]: it was there argued that LD plays a critical role in draining energy from-and thereby establishing the stability of-ITGs in RFPs: within a fluid picture, neglecting LD, RFPs are unstable to ITGs. A kinetic description, with LD included, shows that the same instability is strongly suppressed. Hammett and Perkins [9] devised heuristically a closure procedure to insert approximately kinetic effects (implemented as effective dissipative terms) into fluid codes by matching the kinetic and fluid linear response up to some finite order of an expansion in the small parameter $\omega/(kv_T)$ (low-frequency limit). Later on, this procedure was extended and systematized by several authors. The robustness of the Hammett-Perkins closure has been recently assessed by Sarazin *et al* [10]. In TRB the Hammett-Perkins closure may be turned on or off at will. We will verify, by artificially removing these terms, that instability actually sets in in otherwise stable cases.

The structure of the paper is as follows. In the next section a brief description of the code as implemented for RFX-mod geometry is provided. Then, a selected scan of linear growth rates calculations over some scenarios close to experimental ones will be presented. The aim of this part is essentially to establish consistent conclusions between our results, GS2' and Guo' s ones. i.e., that ITG modes are stable in present-days RFX-mod discharges, although only marginally so in some scenarios. A selected number of cases—of direct relevance for current operating conditions—has been modelled: GS2 turns out to be a more convenient tool if extensive parametric scans are sought [11]. We will be then concerned with addressing the already-mentioned relevance of the LD in establishing the stability of the ITG modes in RFPs. The concluding section summarizes the results as well as provides some discussion concerning the possible role of residual magnetic effects.

## 2. The TRB code and its adaptation to RFP

The TRB code, in the version herein adopted, and for the moment excluding the Hammett-Perkins contribution mentioned before, tracks in cylindrical geometry the 3-dimensional dynamics of five fluid quantities:

$$\begin{aligned}
d_t n_e &= i\omega_{dte}\left(n_{e,0}\phi - p_e\right) + S_n \\
d_t p_e &= i\omega_{dte}\Gamma\left(n_{e,0}\phi + T_{e,0}^2 n_e - 2T_{e,0}p_e\right) + S_{pe} \\
d_t\Omega &= -n_{e,0}\nabla_{||}u_{||} - i\omega_{di}\left(n_{e,0}\phi + p_i\right) - i\omega_{dte}f_t\left(n_{e,0}\phi - p_e\right) + \left[p_{i,0} + f_c n_{e,0}, \nabla_\perp^2\phi\right] \\
d_t u_{||} &= -\nabla_{||}\left(\phi + p_i n_{e,0}^{-1}\right) + S_u \\
d_t p_i &= -i\omega_{di}\Gamma\left[p_{i,0}\left(1 - f_c T_{i,0}/T_{e,0}\right)\phi - T_{i,0}^2 n_e + 2T_{i,0}p_i\right] - \Gamma p_{i,0}\nabla_{||}u_{||} + S_{p_i}
\end{aligned} \tag{1}$$

where $n_e$, $p_{e,i}$ , $\phi$, $u_{||}$ are respectively the electron density, the electron (ion) pressure, the electrostatic potential and the fluid velocity parallel to the magnetic field; $\Omega$ the generalized vorticity = $n_{e,0}\left[f_c\phi/T_{e,0}\right] - \nabla_\perp^2\phi$; $\Gamma$ = 5/3; $f_c = 1 - f_t$. and $f_t$ is the fraction of trapped electrons This latter quantity, in the core, is not noticeably different between Tokamaks and RFPs [12]. The subscripts "||" and "⊥" stand respectively for the directions parallel and orthogonal to the magnetic field. The quantities $S_x$ stand for the source of the quantity $x$. The subscript "0" labels the equilibrium quantitities: in TRB quadratic couplings on the r.h.s. of Eqns. (1) are linearized, and the only nonlinear term retained is the electric drift. Square brackets are the usual Poisson brackets. The time derivative operator is $d_t = \partial_t + [\phi,\bullet] - D\nabla^2$, where $D$ is an artificial diffusivity, introduced both for numerical stability purposes and for modelling background effects (such as collisional effects). Ordinarily, the main scope of $D$ is for avoiding numerical instabilities, and thus is set to very low values: we made selected runs to test the stability of the results against variations, both in value and spatially, of this parameter: our results were insensitive to the changes. All quantities are gyroBohm-normalized (see [7] for a list of normalizations). The symbols $\omega_{dte}, \omega_{di}$ stand for two differential operators, respectively the electron precession drift and the ion magnetic drift. The former writes [13]

$$\omega_{dte} = -i2\varepsilon\left(\frac{1}{4} + \frac{2s}{3}\right)\rho q r^{-1}\partial_\varphi \tag{2}$$

where $s$ is the magnetic shear, $q$ the safety factor, $r$ the radius, ε the inverse aspect ratio, ρ the ion Larmor radius, φ the toroidal angle The definition of $s$ needs to be changed with respect to the standard textbook formula, valid for tokamak geometries, $s = r \times (dq/dr)/q$, which diverges at the reversal surface. A practical adaption to RFPs [5] turns out to be $s = r \times (q/|q|) \times (dq/dr)/\left(q^2 + (r/R)^2\right)^{1/2}$.

The $\omega_{di}$ operator is to be modified when going from tokamak to RFP configuration, since in the latter devices toroidal effects still exist, but are smaller approximately by a factor ε. The operator $\omega_{di}$ is made of two parts: $\omega_{di} = \omega_{d,|B|} + \omega_{d,\mathbf{b}}$. The former term accounts for variations of $|B|$, the latter for changes in its direction.

The two magnetic drift terms above are defined, when applied to an arbitrary function $F$, by

$$\frac{\nabla F \times \mathbf{B}}{B^2} \cdot \frac{2\nabla B}{B} \equiv \omega_{D,|B|} \cdot F$$
$$\frac{\nabla F \times \mathbf{B}}{B^2} \cdot 2\boldsymbol{\kappa} \equiv \omega_{D,b} \cdot F, \quad \boldsymbol{\kappa} = \left(\hat{e}_{||} \cdot \nabla\right)\hat{e}_{||} \quad , \quad \hat{e}_{||} = \frac{\mathbf{B}}{B} \tag{3}$$

The equilibium field is written as

$$\mathbf{B} = \left(0, B_{pol}(r), B_{tor}(r)\left(1 - \frac{r}{R}\cos\theta\right)\right) \tag{4}$$

In the RFP ordering, the poloidal and toroidal components of the magnetic field must be taken as comparable: $B_{pol} \approx B_{tor}$. After some manipulations, one writes to leading order

$$\omega_{d,|B|} + \omega_{d,b} \approx \frac{1}{r^2 B_0^4}\left[r(B_{pol}^2 + B_{tor}^2)' B_{tor} - 2B_{pol}^2 B_{tor}\right] \times \partial_\theta + O(R^{-1}) \tag{5}$$

Where $B_0$ is the value of the field at the core, the prime stands for radial derivative. Eq. (5) restates that toroidal effects (the terms of order $R^{-1}$) are lesser important in RFPs than in tokamaks. For the sake of easiness, we replace in the code the true field profiles with

$$\frac{B_{tor}}{B_0} \approx 1 - r^2, \ \frac{B_{pol}}{B_0} \approx r, \ \frac{R}{L_B} \approx \frac{R}{a} r = \frac{r}{\varepsilon} \tag{6}$$

where $\varepsilon^{-1} \approx 4$ is valid for RFX-mod, and radii are normalized to minor radius $a$

The set of Eqns. (1) is solved by decomposing all functions into linear combinations of basis functions. Poloidal and toroidal dependence is expanded in a Fourier series. Each couple of angular numbers (*m*,*n*), for a given profile of *q(r)*, identifies the location of the corresponding resonant surface. The radial part of the turbulent fields is expressed as a sum of radial wavefunctions centered at the locations of the resonances. Due to the low shear of a RFP a large number of resonant modes is needed in order to provide a full radial coverage (see Fig. 1). By virtue of the choice (6), $B_{\mathrm{tor}}$ reverses only at the very edge. We do not deem this critical since we are interested to the core: TRB is unfit to deal with boundaries because it expands perturbed fields over a set of basis functions with finite support. Whenever the basis functions cross the boundary there is a “spillover” of energy of fluctuating fields, resulting in an artificial severe damping of all perturbations in that region.

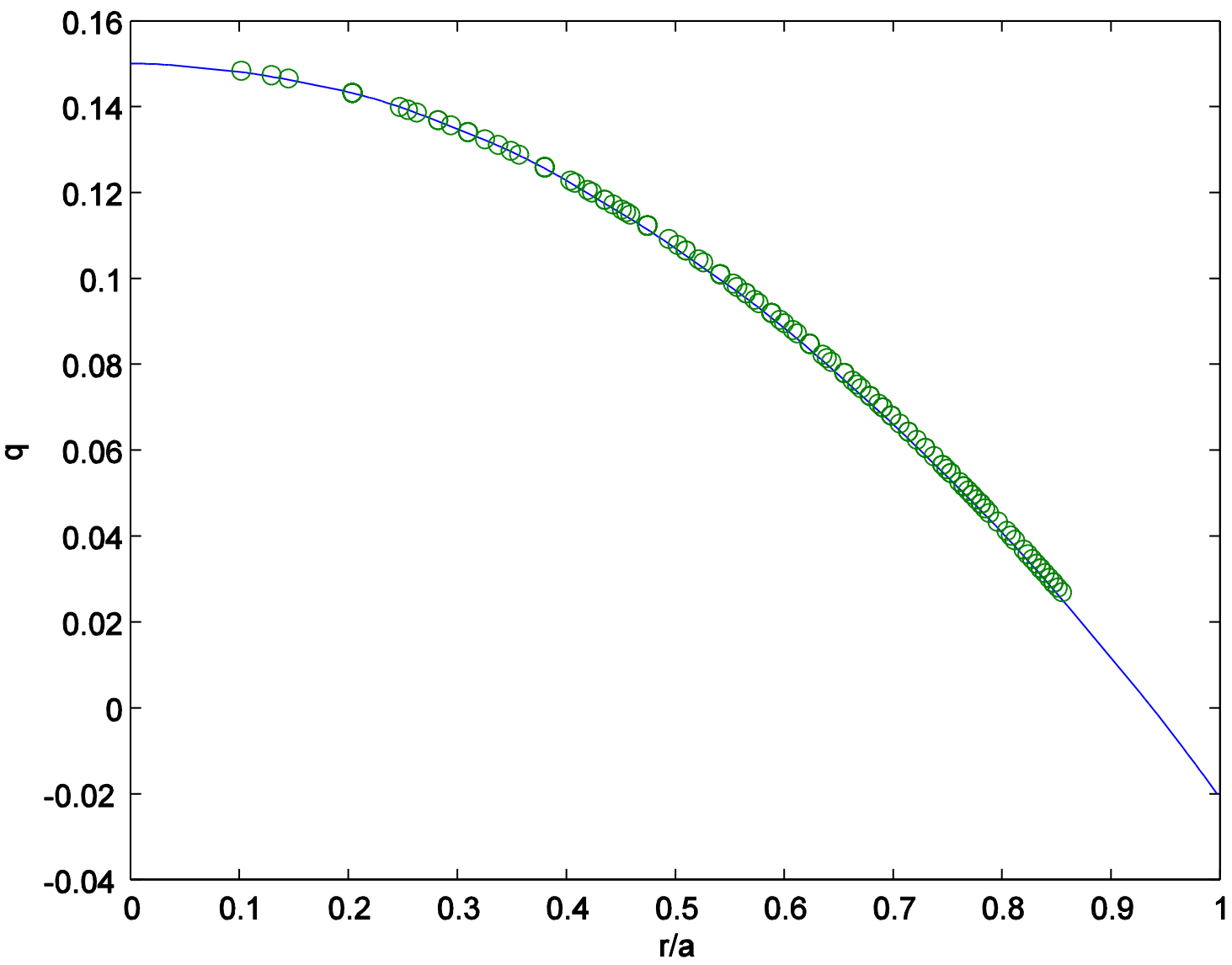

**Fig. 1**. A typical profile for the safety factor $q$ in RFX-mod (solid curve) used in the simulations. Circles tag the location of the modes included in the calculation.

TRB code has been extensively used and tested along years—with success, but in contexts different from ours. It is important therefore to perform comparisons with other approaches whenever possible. The benchmarking between different codes represents a single step within the much broader issue of code validation, which has acquired considerable importance lately. At this stage, we just mention that a few tests run over quite a common set of plasma parameters were performed between TRB and GS2 (GS2, too, was adjusted for RFP geometry, of course). A satisfactory agreement, in term of close guesses for the critical temperature length, was found.

## 3. Selected results

The ITG instability has a threshold behaviour, parameterized by the characteristic temperature length scale $L_T = -T(dr/dT)$, the magnetic shear $s$, and the density length scale $L_{n_e} = -n_e(dr/dn_e)$. It is not the aim of this paper to present a full scan over these parameters, which is covered by the companion work [11]. Our scope is to restrict to a rather limited set close to realistic values or to values that could become realistic in the near future if the current trend of performance increasing should continue.

Thus, the first step is assessing what the most important parameters for the scan are. In RFX-mod, density profiles are usually fairly flat or even slightly hollow, since particle fuelling occurs only at the edge through desorption from the walls. Hence, from now on, we will keep density profile fixed and flat $(L_{n_e} = \infty)$. Conversely, heating is just ohmic and hence heat source is spred everywhere. Electron temperature is routinely measured through RFX-mod Thomson scattering [14]. The relevant quantity to our purposes is the ion temperature, which is not—to date—measured on RFX-mod. There exists, though, indirect evidence from spectroscopic Doppler measurements of the temperature of partially ionized impurities, suggesting a likely value of the ratio: $0.5 < T_i/T_e < 0.8$ [4]. No direct measurements of the magnetic shear exist in RFX-mod as well. The profiles of $q$ and $s$ are reconstructed from numerical models for equilibrium profile, and hence subject to some uncertainty: see [15] for an example of the reconstruction of the profile of $q$. In the following we will use a simple second-order polynomial for $q$, as shown in Fig. (1). In conclusion, the temperature profile, and hence $L_T$—although with the above caveat concerning the uncertainties about it—still remains the most convenient parameter upon which performing a scan, in particular in view of the fact that temperature is the most natural variable quantifying machine performances.

Fig. (2) summarizes our simulation. The uppermost plot shows the ion temperature profiles adopted for our simulations. Notice that the normalization used are such that Temperature = 1 in dimensionless form means 1 keV in physical units. Currently, electron temperatures in present-days QHS RFX-mod operations range around—and somewhat above—1 keV [4] (while are slightly more than one half in MH operations [16]) The magenta curve in Fig. (2) should be considered a plausible approximation to present-day RFX-mod ion temperature profiles; basically, it is a rescaled electron-temperature-profile: compare it, e.g., with Fig. 5 of ref. [5]. It features a central plateau and a rather gentle falling towards the edge. The remaining curves are tailored in order to study parametrically the effects of a sharp temperature gradient in the middle of the radius. The results are provided in the plot below, in terms of the maximum growth rate at each spatial point, with positive values flagging instability. The third plot quantifies the logarithmic derivative $L_T = -T(dr/dT)$ for all the profiles. It is qualitatively clear that the onset of instability is close to $L_T \approx 10$ cm. This is done more quantitative in the last plot, where we have plotted the growth rate γ versus $L_T$ (both evaluated at about the location of the maximum for γ). The points accommodate approximately along a straight line that intercepts the horizontal axis at $L_T^{(\mathrm{crit})} \approx 10.5$ cm.

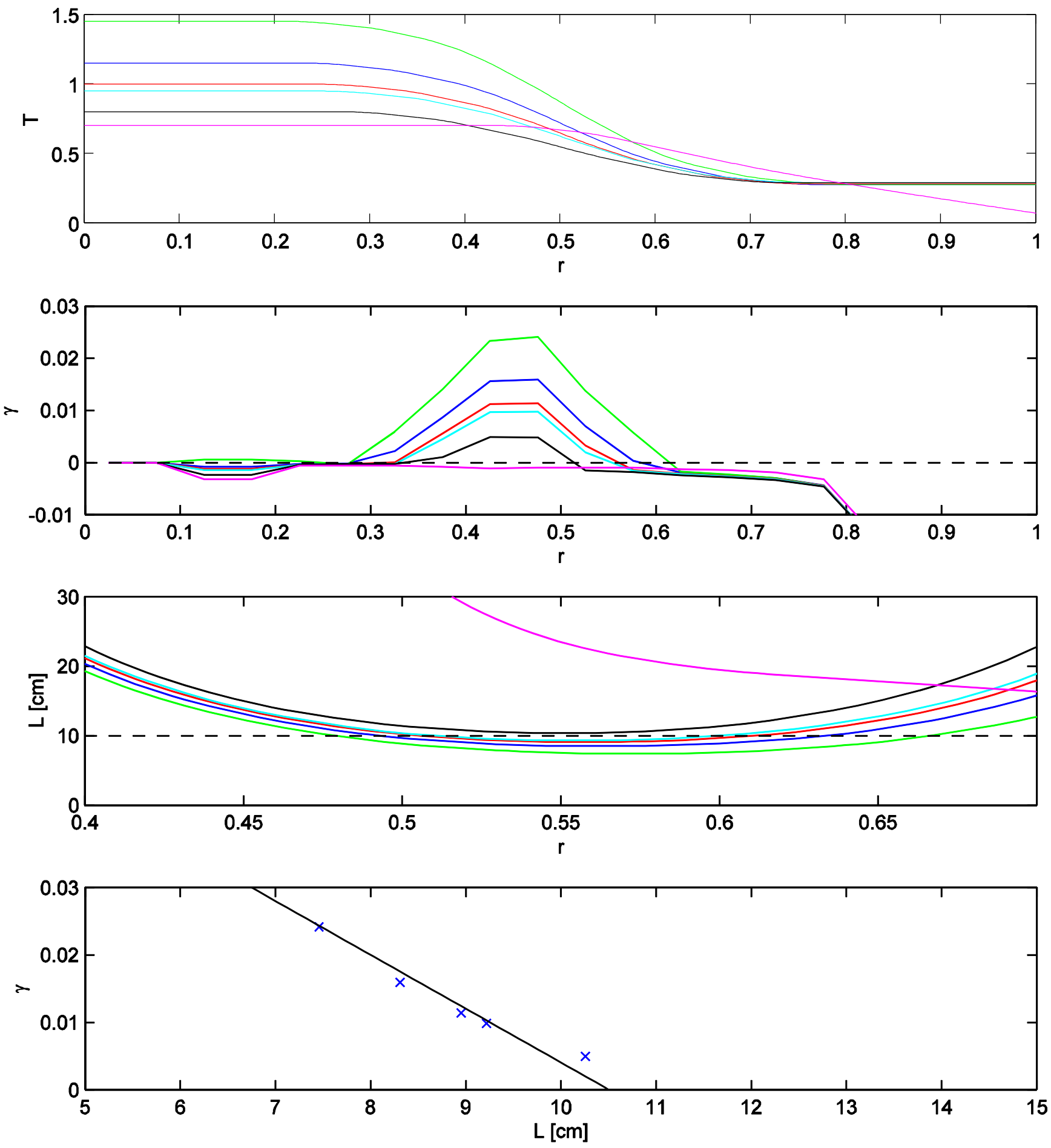

**Fig. 2**. (Color online) From top to bottom: ion temperature profiles used in the simulations; local maximum growth rate; ion temperature logarithm length scale; maximum growth rate versus length scale with superimposed a straight line. Electron temperature was scaled correspondingly by a factor 1.25, *i.e*, $T_i / T_e \approx 0.8$ .

In Fig. (3), we give additional evidence in favour of the ITG character of the unstable modes shown in Fig. (2).

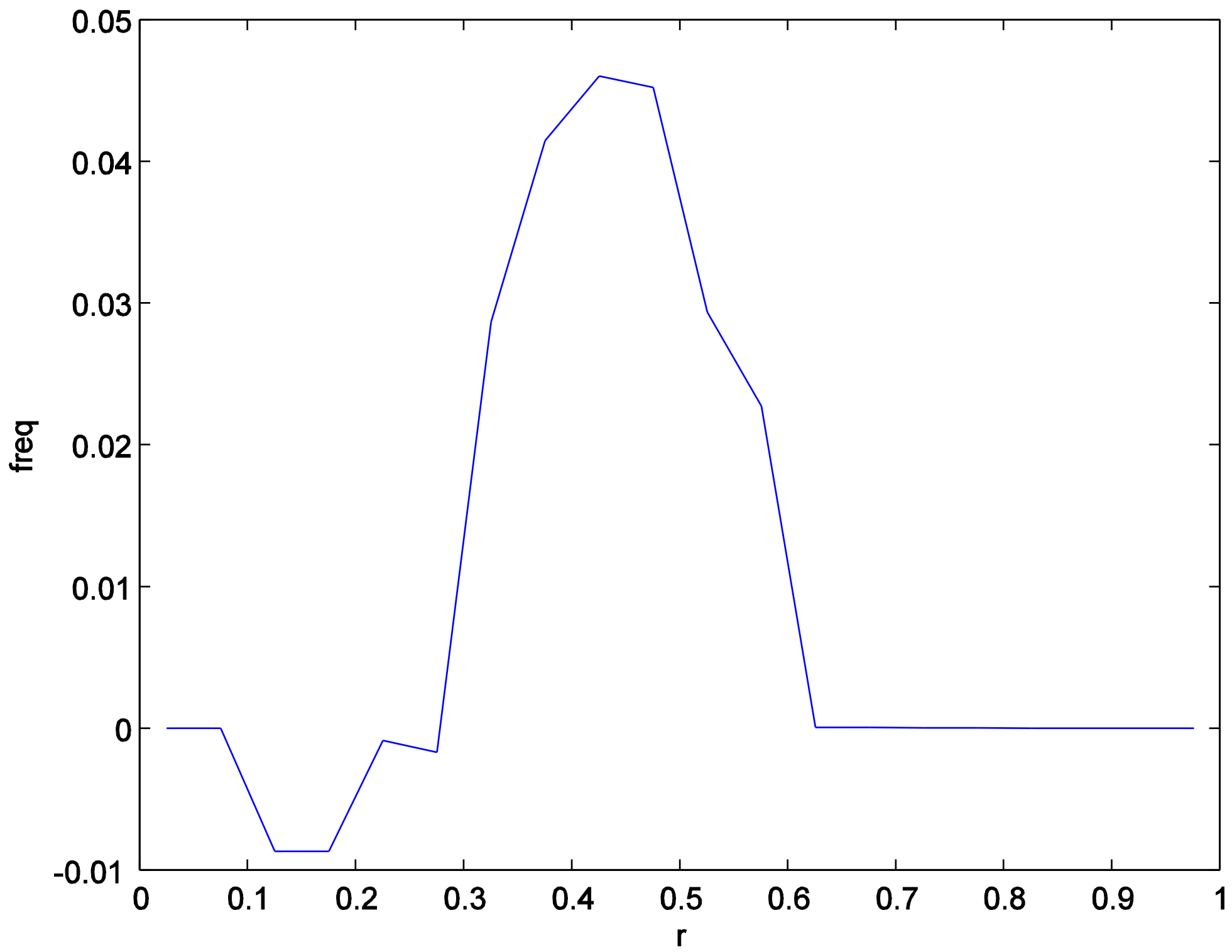

**Fig. 3**. Typical mode rotation frequency for the cases displayed in fig. (2). The positive sign means directed along the ion diamagnetic direction, the expected direction of propagation for ITGs.

How does our result compare against experiment ? We have attempted to provide a rough answer to this question, by taking advantage of the several SHAx temperature profiles published in Bonomo *et al* [17]. The logarithmic length scale has been evaluated very roughly from published figures as $L_T^{(\exp)} \approx (x_{low} - x_{high}) / \ln(T_{high} / T_{low})$, where $T_{\mathrm{high}}$ is the temperature measured at the edge of the high-temperature SHAx region, $T_{\mathrm{low}}$ the temperature at the foot of the linear slope region (in all cases the outboard semiaxis has been considered), and $x_{\mathrm{in,out}}$ the corresponding radii. Here, we are still dealing with *electron* temperature, while we recalled earlier that *ion* temperature is likely a factor at least 80% lower in the core. Nearer to the edge, on the other hand, due to higher collisionality, the ratio $T_i/T_e$ should be closer to unity. As a further exercise, therefore, we recomputed the same $L_T^{(\exp)}$, but with reduced $T_{\text{high}}$ : $T_{\text{high}} \rightarrow T_{\text{high}} \times 0.8$, whereas left fixed $T_{\text{low}}$ . Results are summarized in Table 1.

| Shot | $L_T^{(\exp)}$[cm] | $L_T^{(\exp)}(\times 0.8)$ [cm] |
|---|---|---|
| 24932 | 18 | 25 |
| 23977 | 14 | 17 |
| 23912 | 13 | 17 |
| 24598 | 13 | 17 |

**Table 1.** Middle column, electron temperature logarithmic length scale for the shots listed on the left column. Right column, the same quantity evaluated using "rescaled" temperature, to account for ionic profiles, as explained in the text. For completeness, although it should be of no relevance in the present context, we mention that discharges 23977 and 23912 featured OPCD.

The likely logarithmic length scale is larger than the critical one, hence this analysis suggests that present-days RFX-mod profiles are sub-critical—and hence stable—to ITG modes. The exact distance from criticality, quantified as $(L_T - L_T^{crit})/L_T^{crit}$, is unknown but it is not likely to be large: less than 30% according to table 1, in the most "favourable" case. It cannot be therefore discarded the possibility that supercritical situations be transiently and locally attained, in consequence of large temperature fluctuations.

.

**4. On the effect of Landau Damping**

In this section we turn to address briefly the intertwining between LD and the magnetic structure[2]. We propose a comparison of growth rates with—and—without the LD effect. Two choices for the safety factor are compared: one "RFP-like", with $q$ small (<<1) and monotonously decreasing, and another "tokamak-like", with $q > 1$ and increasing (see Fig. 4). Temperature profile is shown in the same figure, while density is held flat.

[2] Recalling again that we can deal only with a (close) surrogate of the full LD modelization.

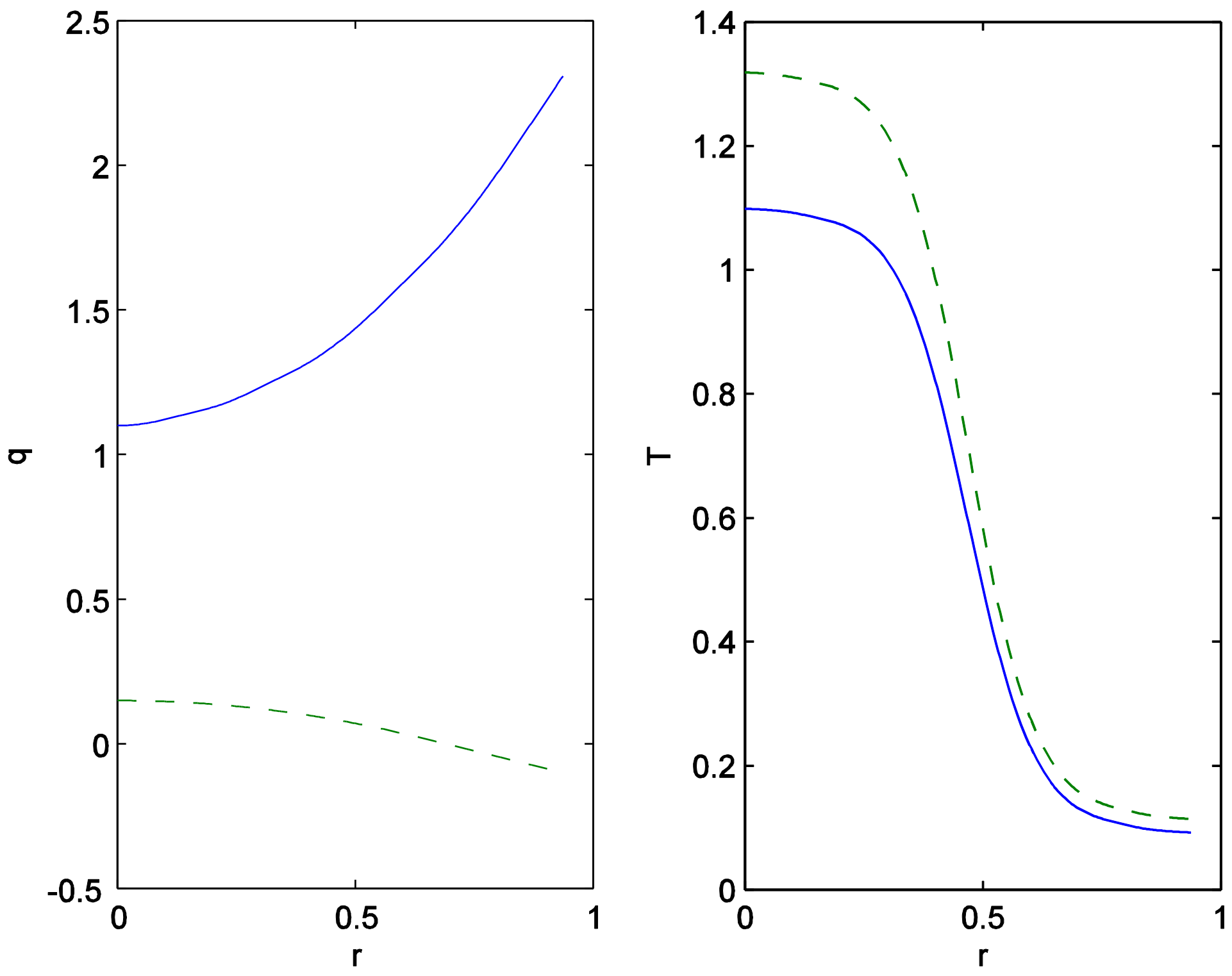

**Fig. 4**. Left plot: dashed curve, the RFP-like $q$ profile, similar to Fig. (1); solid curve, the tokamak-like profile. Right plot: ion temperature (solid curve) and electron one (dashed curve).

We have chosen a case that—in the RFP configuration—is ITG-unstable even with LD turned on, in order to assess the degree of stabilization brought by the mechanism. Indeed, see Fig. (5), LD almost halves the growth rate in the RFP case, while its contribution is much more modest (order 15%) in the "tokamak" case.

One natural question is wondering to what extent results from TRB do depend upon the choice of the closure. We have advanced in the Introduction some arguments supporting the view that Hammett-Perkins' as rough as it can be, looks accurate enough for our purposes. Present results, confirming independent gyrokinetic simulations, further strengthen this position.

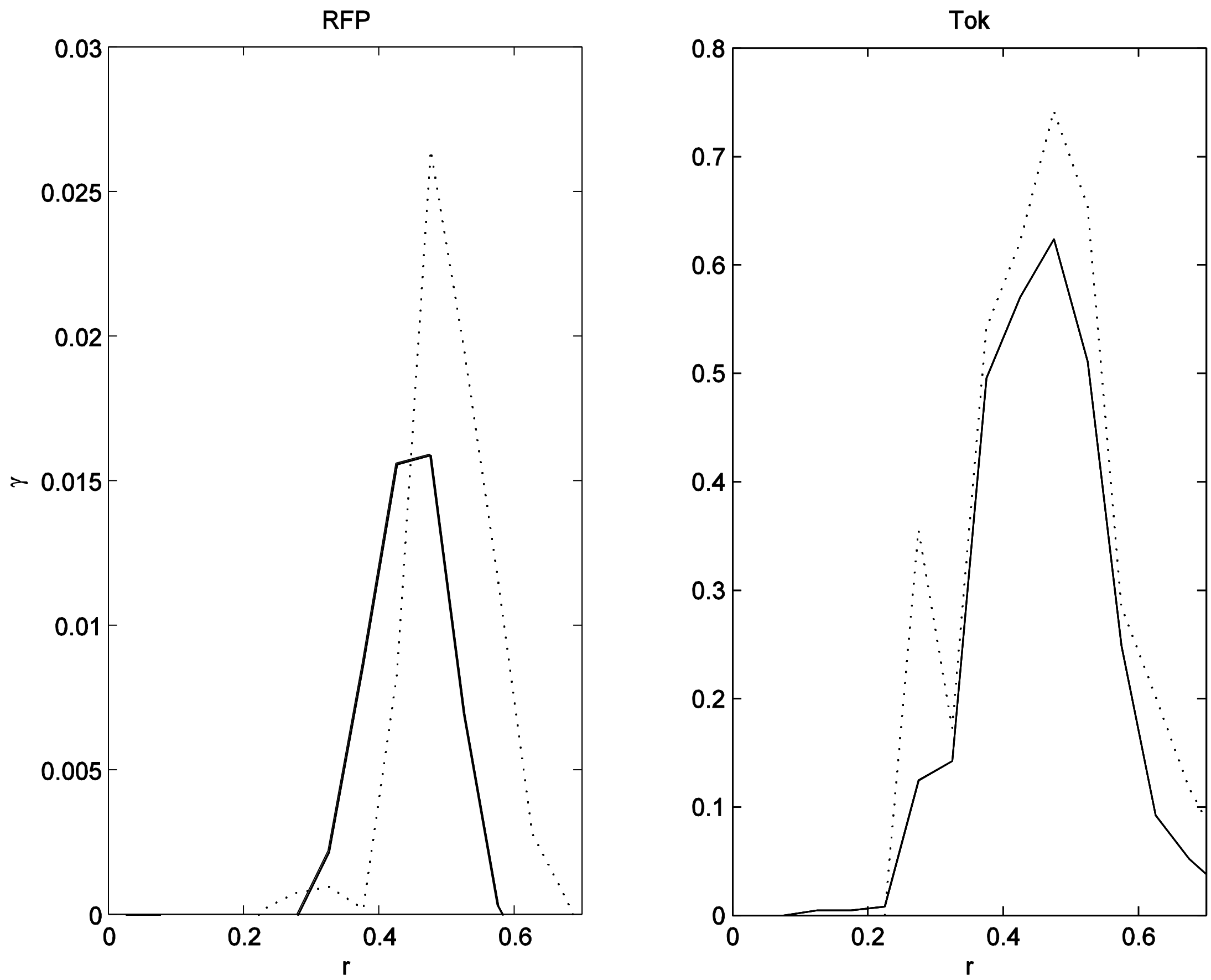

Fig. **5**. Maximum growth rate *versus* radius (positive values meaning instability). Solid curve, result from a TRB run including Landau damping; dotted curve, Landau damping has been turned off. Left plot, RFP-like case; right plot, tokamak-like case. Plasma conditions are those shown in Fig. 4.

**5. Final discussion**

TRB simulations point it out that, due to kinetic effects, present-days RFX-mod temperature profiles are stable to ITG turbulence, not dramatically far from marginal stability though. ITG turbulence might thus become an issue in the future should current improvements towards the buildup of stronger and sharper internal transport barriers continue.

Our results, together with earlier gyrokinetic approaches, papers [5,11], form a consistent body of evidence[3]. We think that this concurring approach (kind-of benchmarking) be quite valuable in the light of the present-days interest towards the broader issue of code validation [19]. Finally, as an important aside, consistency of our findings with those from gyrokinetic codes strengthens the confidence in the

[3] We mention also one recent study that tackles this subject with even more detail, through a kinetic integral approach with consideration of finite Larmor radius effects [18].

Hammett-Perkins closure approximation as a reliable way to incorporate kinetic effects into fluid formulations.

We have not here treated the Trapped Electron Modes, that develop at the same range of wavelengths as ITGs and are therefore commonly treated on equal footing. Studies on this instability are still quite preliminary and will be presented in further publications; first results but suggest that these modes could be relevant in RFPs in some peaked-density scenarios [20].

An issue that has been deliberately postponed up to this point concerns the coexistence and the possible mutual interactions of electrostatic and magnetic effects. From the one hand, magnetic effects may act addictively, triggering further instabilities. This turns out to be important in connection with the question: What does limit the growth of core temperature and related gradients in the absence of ITG/TEM turbulence? We remind that there is wide consensus about a severe reduction of magnetic chaos within the investigated region. However, some residual magnetic turbulence may still survive. It may come from large-scale MHD tearing modes, but a promising line of investigation concerns the possible role of microscopic electromagnetic modes: microtearing modes. A study is currently being carried out using GS2 about their linear stability, and its results suggest quite convincingly that, under ordinary SHAx conditions, microtearing modes are indeed unstable. One point that needs clarification is the level of transport that they may drive; however, quasi-linear estimates suggest that the order of magnitude is the right one to match power-balance estimates [21]. Thus, it may be that ultimately (at least part of) the needed transport is provided by tearing turbulence—both large- and small-scale.

On a different aside, the coupling between the scalar potential and the vector potential—may potentially modify the nature of otherwise purely electrostatic modes: in the presence of both a scalar potential perturbation $\phi$ and a magnetic vector potential $A_{||}$, the full potential exerted on particles is $\phi' = \phi - v \times A_{||}$. This kind of coupling is commonly neglected in Tokamaks and—of course—it could not *a priori* be accounted for in the present study, since TRB can deal only fluid electrostatic fluctuations. Investigations using GS2 are presently being planned.

## Acknowledgments

I. Predebon is acknowledged for the simulations done with GS2 code. F.S. wishes to thank L. Garzotti for help with TRB code. S. Cappello and D. Escande encouraged and helped with this work. A. Alfier spotted several inaccuracies. This work was supported by the European Communities under the contract of Association between EURATOM/ENEA. The views and opinions expressed herein do not necessarily reflect those of the European Commission.